\documentclass[aps,pre,showpacs,twocolumn]{revtex4-1}

\usepackage{graphicx}
\usepackage{latexsym}

\newcommand{\Tilde}{\raise.17ex\hbox{$\scriptstyle\sim$}}
\begin{document}

%Title of paper
\title{Opinion dynamics of random-walking agents on a lattice}

\author{Suhan Ree}
\email{suhan@physics.utexas.edu}
\affiliation{Department of Industrial Information, Kongju National University, Yesan-Up, Yesan-Gun, Chungnam, 340-702, South Korea}	
\affiliation{Center for Complex Quantum Systems and Department of Physics, University of Texas at Austin, Austin, TX, 78712, USA}

\date{\today}

\begin{abstract}
Opinion dynamics of random-walking agents on finite two-dimensional lattices
is studied.
In the model, the opinion is continuous,
and both the lattice and the opinion can be either periodic or non-periodic.
At each time step, all agents move randomly on the lattice, and 
update their opinions based on those of neighbors with whom
the differences of opinions are not greater than a given threshold.
Due to the effect of repeated averaging, opinions first
converge locally, and eventually reach steady states.
Like other models with bounded confidence, steady states in general are
those with one or more opinion groups, 
in which all agents have the same opinion.
When both the lattice and the opinion are periodic, however, 
metastable states, in which
the whole spectrum of location-dependent 
opinions can coexist, can emerge.
This result shows that, when a set of continuous opinions forms a structure like a circle,
other that a typically-used linear opinions, rich dynamic behavior can arise.
When there are geographical restrictions in reality, 
a complete consensus is rarely reached, and 
metastable states here can be one of the explanations for these situations, especially
when opinions are not linear.
\end{abstract}

% 89.75.Fb : Structures and organization in complex systems
% 87.23.Ge : Dynamics of social systems
% 02.50.Ey : Stochastic process
% 05.40.Fb : Random walks and Levy flights


\maketitle

% ====<Introduction>====================
\section{Introduction}
The attempt to investigate social systems by physicists is several decades old\cite{Galam:1982},
even though social dynamics has become a popular subject in statistical physics only recently\cite{Castellano:2009}.
One of the reasons for this interest is that one basic approach to study social systems is similar to what
statistical physicists typically try to do:
namely, finding macroscopic behavior or emergence from dynamics of microscopic entities\cite{Schelling,Liggett}.
While physical systems deal with particles, 
entities that make up social systems are humans, or groups of humans.
Figuring out dynamic behavior of even one human being is not an easy task,
but some aspects of collective behavior of many individuals
are known to be describable using microscopic models\cite{Schelling}, 
and even universal\cite{Fortunato:2007}.
%which makes this approach meaningful.
In addition, due to the current ubiquity of the Internet, especially the popularity of social networks,
and the increased capability of processing vast amount of social data, 
this kind of approach has become not only possible, but also useful.

Opinion dynamics is one of the social-dynamics problems that can be closely related to physical problems.
%and physicists use simple microscopic models to describe such problems.
%that can explain everyday experiences on opinion changes in a society.
Microscopic models we are interested in here typically evolve with discrete time steps, and have 
the fixed number of ``agents" (actors, or individuals)  with their own opinions.
We can categorize these models using several basic features.
Opinions can be discrete\cite{deOliveira:1992,Galam:1986,*Galam:2004,*Galam:2005a,*Galam:2005b,Galam:1998,
SznajdWeron:2000,Ianni:2002,Vazquez:2003,BenNaim:2003,
Gil:2006,Fu:2008,Mandra:2009,Benczik:2009},
or continuous\cite{Chatterjee:1977,Hegselmann:2002,
Weisbuch:2002,BenNaim:2003,Weisbuch:2005,
Fortunato:2005,Holme:2006,Lorenz:2007,Kozma:2008,Iniguez:2009}.
Examples of discrete opinions are yes or no on a question (2 values),
evaluation on a scale from 1 to 5 (5 values), 
choices in elections (2 or more values), and so on.
When there are more than a few choices, however, continuous opinions
can be used:
fine-scaled evaluation on something on a scale from 0 to 1,
political views, and so on.
An opinion can be a vector of integers\cite{Axelrod:1997,Weisbuch:2002,Laguna:2003,Vazquez:2007} as well.
Another important feature is how a model restricts interacting partners of an agent at a given time.
Agents typically have ongoing relationships with others, and interact with selected peers out of related ones.
The structure of these relationships plays an important role in social dynamics,
and networks can be used to describe these relations (social networks).
We can divide models into three different cases:
(i) fully-connected networks, where
each agent is related to all other agent at any moment (there is
no restriction, and the network concept is not necessary)
\cite{Deffuant:2000,Hegselmann:2002,Weisbuch:2002,
BenNaim:2003,Weisbuch:2005,Laguna:2003,Vazquez:2007};
(ii) fixed networks, where each agent is related to the limited number of agents given by time-independent 
networks
\cite{deOliveira:1992,Axelrod:1997,SznajdWeron:2000,Ianni:2002,Laguna:2003,Vazquez:2003,Weisbuch:2002,Weisbuch:2005,
Fortunato:2005};
(iii) evolving networks, where
network structures evolve with time
\cite{Chatterjee:1977,Gil:2006,Holme:2006,Fu:2008,Gross:2008,Kozma:2008,Iniguez:2009,Mandra:2009,Benczik:2009}.
Finally, models can be differentiated by how updating agents are chosen at each time step.
One can update one agent at a time (the {\em serial} update),
or all agents synchronously, especially when the order of update doesn't play a role 
(the {\em parallel} update)\cite{Hegselmann:2002,Fortunato:2005}.

The model introduced here uses continuous opinions with evolving networks,
and the parallel update. 
Agents reside and move randomly on a two-dimensional (2D) lattice.
At each time step, agents update their locations in the lattice using the
2D random walk, and change their opinions synchronously.
Only nearest-neighbor interactions are allowed for opinion changes; hence
interacting partners can be represented by a
{\em contact network}\cite{Stehle:2010}, whose evolution is 
only governed by the movements of agents.
This is basically the process of repeated averaging\cite{Feller,Chatterjee:1977}, 
and opinions have a tendency to move toward those of neighbors.
Our model also uses the threshold to restrict interactions between agents with the big difference of opinions
({\em bounded confidence}, and
the use can be justified in many aspects\cite{Festinger,Gal:2010}).
In most models with the above setup
\cite{Deffuant:2000,Hegselmann:2002,Weisbuch:2002,Vazquez:2003,
BenNaim:2003,Amblard:2004,Weisbuch:2005,Fortunato:2005,Lorenz:2007,Kozma:2008},
the system eventually reaches one of steady states, where
one or more groups of agents reach their consensus.
Unlike other models, we assume that both the lattice and the opinion can be periodic.
The shape of the lattice can be either rectangular or toroidal:
two of the simplest shapes, and yet different topologically.
Opinions can be periodic, too, when 
an opinion is about a periodic subject like the time of the year.
When both the lattice and the opinions are periodic, we observe some periodic
{\em metastable} states.
In these states, there is no consensus even though
opinions converge locally, 
and the whole spectrum of opinions, which depend only
on spatial locations of agents, can coexist.
The main purpose of this work is to 

The plan of this paper is as follows.
In Sec.~II, the model is described, and steady states of an extreme case of an one-dimensional lattice are found.
In Sec.~III, numerical results when both the lattice and the opinion are periodic are shown.
Finally, in Sec.~IV, possible extensions, and many aspects of this model are discussed.

%=====<Model>======================
\section{Model}
\begin{figure}
\includegraphics[scale=0.8]{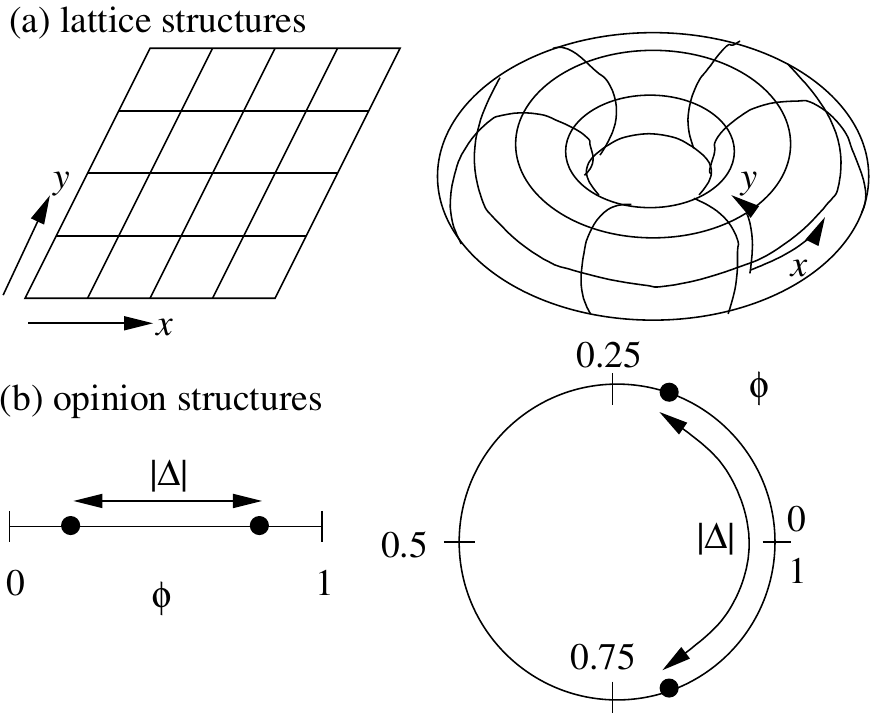}
\caption{\label{topology}
 	Structures of the lattice and the opinion. 
	Note that, for example, the differences of opinions between $0.2$ and $0.8$ are different for two cases:
	0.6 and 0.4, respectively. 
	}
\end{figure}

We propose a simple model for opinion dynamics of random-walking agents with only nearest-neighbor interactions.
%To make the model more realistic, 
We assume that there are $N$ agents, each with an opinion, and that they reside on finite 2D lattices.  
Opinion changes will come only from interactions with neighbors.
Time is discrete, and is represented by a dimensionless quantity, $t$, 
which is a non-negative integer.
Agent $i$ ($1\le i\le N$) will have a location and an opinion
at time, $t$.
Because the structure of the lattice can play an important role in the dynamics,
we consider two structures: a rectangle (non-periodic) and a 2D torus (periodic)
[see Fig.\ \ref{topology}(a)].
%==== figure <topology>
%====
The location on the lattice  for agent $i$ at time $t$ is
($x_i, y_i$), where $x_i$ and $y_i$ are non-negative integers
($0\le x_i < X$ and $0\le y_i < Y$).

In our model, an opinion of agent $i$, $\phi_i$, is a real number between 0 and 1 ($\phi_i\in [0,1]$),
and can be either periodic or non-periodic [see Fig.\ \ref{topology}(b)].
For non-periodic opinions, $\phi_i$ is a number on a line between 0 and 1;  while,
for periodic opinions, $\phi_i$ is a point on a circle (for this case, $0$ and $1$ are the same opinion).
Then, the state of agent $i$ at time $t$, $s_i(t)$, is represented by three numbers,
\begin{equation}
s_i(t)=[x_i(t),y_i(t);\phi_i(t)],
\end{equation} 
and the state of $N$ agents is an $N$-tuple of $s_i$'s,
\begin{equation}
s=(s_1,s_2,\ldots,s_N).
\end{equation} 
%The state space of $s_i$ is $L\times O$, where $L$ represents the space for a 2D lattice,
%and $O$ represents the space for opinions; hence the state space for the whole system
%becomes $S^N=L^N\times O^N$.

How do states of agents evolve?
The change of $s_i$  will depend on states of other agents.
For each agent, there are two kinds of movements: changing locations in the lattice, and change of opinions.
First, we look at how agents move in lattices.
For simplicity,  we use an independent random-walk motion for each agent.
The movement of agent $i$ ($1\le i\le N$) is represented by possible choices of locations
at a next time step as follows,
\begin{equation}
(x_i, y_i)\rightarrow \left\{
	\begin{array}{l}
		(x_i+1,y_i)\\
		(x_i,y_i+1)\\
		(x_i-1,y_i)\\
		(x_i,y_i-1)\\	
		(x_i,y_i)
	\end{array}\right. .
	\label{movement}
\end{equation}
Here an agent can move to five neighboring locations, including an option for staying\footnote{
%The main reason to include the previous location in the possible destinations 
%in the next step is to make the state space for locations of all agents, $L^N$, irreducible
If $X$ and $Y$ are even numbers,  some agents will not meet at the same
location forever if only four moving choices are given in Eq.~(\ref{movement}).}
%In other words, $L^N$ is divided into two equivalent classes of states,
%and, if the interaction only
%happens among the agents at the same location, 
%agents are also divided into two groups 
%in which only agents in the same group can interact.}, 
with equal probabilities, 1/5. 
If an agent has less than five choices to move, 
it will move to one of those locations with equal probabilities as well.
For the rectangular lattice, for example, if an agent is on an edge,
it has four possible choices with equal probabilities, 1/4.
We assume that more than one agent can reside in the same location, and
we define neighbors as agents residing at the same location 
at a given time.
Since agents move randomly, neighbors of an agent will also change with time.
This contact network consists of disconnected cliques of various sizes 
as in Ref.~\onlinecite{Stehle:2010}.

Interacting partners of an agent are further reduced by the use of the threshold $d$
as in other models of bounded confidence\cite{Deffuant:2000,Weisbuch:2002,Weisbuch:2005,Hegselmann:2002,BenNaim:2003,
Vazquez:2003,Amblard:2004,Fortunato:2005,Weisbuch:2005,Kozma:2008,Lorenz:2007}.
%Opinions will change as a result of interactions with neighbors.
%Our rule for opinion changes has two basic assumptions: 
%(i) agents are influenced by neighbors with similar opinions only (bounded confidence), and (ii) 
%an opinion of an agent will move toward those of neighbors in a deterministic way.
%These assumptions can be realized in many ways, and we modify the Deffuant model\cite{dw}.
%To realize the first assumption, a threshold, $d$, is used.
If the difference of opinions for a given pair of agents is greater than $d$
there will be no influence.
In other words, if we introduce 
$\Delta_{ji}$ as the difference of opinions between agents $i$ and $j$,
agent $i$ will not be influenced by agent $j$ if $|\Delta_{ji}|$ is greater than $d$.
For non-periodic opinions, $\Delta_{ji}$ can be obtained by subtraction,
\begin{equation}
	\Delta_{ji}=\phi_j-\phi_i.\label{linear}
\end{equation}
For periodic opinions, on the other hand, we set
\begin{equation}
	\Delta_{ji}=\left\{
		\begin{array}{ll}
			\phi_j-\phi_i+1 & (\mbox{if}\ \phi_j-\phi_i\le-0.5),\\
			\phi_j-\phi_i-1 & (\mbox{if}\ \phi_j-\phi_i >0.5),\\
			\phi_j-\phi_i    & (\mbox{otherwise}),
		\end{array}\right.
	\label{periodic}
\end{equation}
[see Fig.\ \ref{topology}(b)].
The ranges of both $\Delta_{ji}$ and $d$ are different for two types of opinions: for non-periodic opinions, 
$-1\le \Delta_{ji}\le 1$ and $0<d\le1$, and for periodic opinions, $-0.5<\Delta_{ji}\le 0.5$
and $0<d<0.5$ 
($d=0.5$ for periodic opinions is excluded 
because an uncertainty can be introduced.)

Then, we can define the set of neighbors with similar opinions as $K_i$,
where agents in this set are neighbors of agent $i$ at a given time
and absolute values of differences of their opinions with agent $i$ are
less than or equal to $d$.
By including agent $i$ in $K_i$ ($i\in K_i$), there will be at least one element in $K_i$.
Then, the opinion of agent $i$, $\phi_i$, at $t+1$ becomes
\begin{equation}
% \phi_i \rightarrow \left\{
% 	\begin{array}{ll}
% 		\phi_i+g\frac{\sum_{j\in K_i}\Delta_{ij}}{|K_i|} & (\mbox{if}\ |K_i|>0), \\
% 		\phi_i & (\mbox{if}\ |K_i|=0).
% 	\end{array}\right.
\phi_i(t+1) = \phi_i(t)+g\frac{\sum_{j\in K_i(t)}\Delta_{ji}(t)}{|K_i(t)|}, 
\label{opinion_change}
\end{equation}
where $g$ is a convergence parameter ($0<g\le1$),
and $|K_i(t)|$ is the number of elements in the set $K_i$ at time $t$.
When $|K_i|=1$, there is no interaction for agent $i$, and $\phi_i$ will not change.
When $|K_i|=2$, the interaction is binary, and Eq.~(\ref{opinion_change}) 
becomes the equation in 
the Deffuant model\cite{Deffuant:2000}
(in the original model, $\mu$ has been used, which is basically $g/2$).
When $|K_i|>2$, agent $i$ is interacting with more than one neighbors at once as in the Hegselmann-Krause 
model\cite{Hegselmann:2002}.
%The model can be modified to allow
%only binary interactions. 
%For example, an agent can choose one interacting counterpart randomly
%out of $K_i$, and
%There will not be much difference to the results, especially when the density
%of agents is not big.

The right-hand side of Eq.\ (\ref{opinion_change}) can be also written as
$\phi_i+g(\bar{\phi_i}-\phi_i)$, where $\bar{\phi_i}$ is the average opinion of agents in $K_i$.
When $g=1$ and $d$ is maximal,
all interacting agents
will have the same averaged opinion at the next step. For example, for non-periodic opinions, when two agents with $\phi_1$ 
and $\phi_2$ interact, both of their opinions will become $(\phi_1+\phi_2)/2$ at the next step.
Note that care has to be taken when the opinion is periodic. For example, the average of opinions, 0.1 and 0.9,
should be 0, not 0.5 as in the non-periodic case.
For periodic opinions, the value of $\phi_i$ can become greater than 1 or less than 0
after Eq.\ (\ref{opinion_change}) is applied;
in those cases, we can adjust $\phi_i$ value by subtracting or adding 1 to keep
$\phi_i$ in [0,1].
When $0<g<1$, every interacting agent
will move toward a certain value at the next step.
%(if all neighbors' opinions are within
%the range of $d$, this value is the average opinion of all neighbors). 
If the same agents interact for
more than one time step, their opinions will gradually converge to one opinion value.
%(they will not converge when $d$ is smaller than the difference between
%any pair of neighbors).
The bigger the value of $g$, the faster they will converge.
We are not considering the case with $g=0$ or $d=0$, because the model becomes trivial.
This model has five parameters: $N$, $X$, $Y$, $d$, and $g$.
%For non-periodic opinions, the total opinions will not be conserved unless $d=1$.
%If only binary interactions
%are allowed, which is not the case for our model according to Eq.\ (\ref{opinion_change}),
%opinions will be conserved for any $d$.
%On the other hand, for periodic opinions, opinions are not conserved in any case.

At $t=0$, the state of all agents is given, and 
opinions and locations thereafter will be calculated from the state of the previous time step.
At each time step, agents undergo random walks on the lattice, and
change opinions according to Eq.\ (\ref{opinion_change}) synchronously.
This dynamic process is stochastic because of random walks, even
though opinion changes are deterministic.
This also is a Markov process 
because the state at the previous time step is all we need
to find the current state and beyond.

\begin{figure}
\includegraphics[scale=0.45]{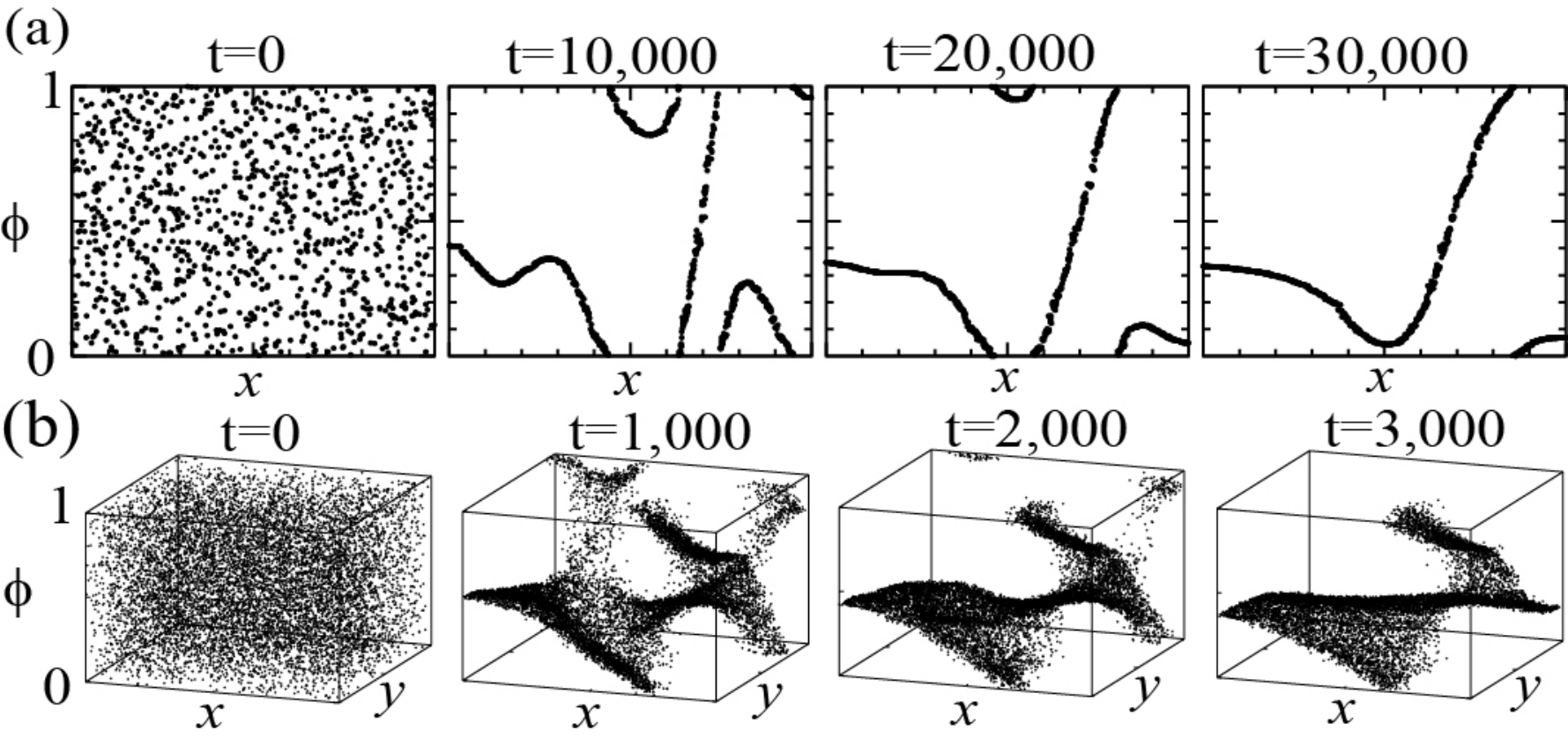}
\caption{\label{evolution}
	Scatter plots to show time evolutions in 1D and 2D cases 
	with the non-periodic lattice and periodic opinions.
%	Only initial states and transient states are shown here.
%	We can observe that states are quickly compressed to 1D and 2D 
%	surfaces, respectively in both cases.
	(a) 1D case: $N=1000$, $X=1000$, $Y=1$, $g=1$. $d=0.49$; 
	(b) 2D case: $N=10000$, $X=100$, $Y=100$, $g=1$, and $d=0.49$. 
	}
\end{figure}

Before looking at numerical results, we can
get some insights by looking at an extreme case:
the one-dimensional (1D) case (by setting $Y$ as 1) with $g=1$ and the maximal $d$. 
In this case, opinions of agents at one location converge to the same value at each time step
according to Eq.~(\ref{opinion_change}), and due to local interactions,
the whole state can be approximated by a continuous 1D curve in ($x$, $\phi$)-space
[see Fig.\ \ref{evolution}(a)].
%==== figure <evolution>
%====
Then, we can focus only on dynamics of 
this curve instead of $s$ in Eq.\ (2).
The local stability at a location can be achieved if the curve is locally linear
(see Appendix for detail), and we get the global stability when the curve is linear 
everywhere, which means that steady states are straight lines in $(x,\phi)$.
Given an initial condition, the system will eventually reach one of  
states represented by straight lines (it can  be seen as a straightening process of a curve), 
and opinion values with respect to $x$ will not change.
The most common is the one where every agent has the same
opinion (complete consensus), as most models of opinion dynamics have found.
Once reached, opinions of all agents will not change afterward.
We will call these steady states {\em flat} here, because they
can be represented by flat lines in $(x,\phi)$.
If we use the term an ``opinion group" (a cluster, or a party) as a set of agents 
that have reached a consensus,
there will be only one opinion group for maximal $d$, while there can be more than one 
opinion groups when $d$ is small (say, $d<0.3$), as seen in models of bounded confidence.

When both the lattice and the opinion are periodic, however, {\em non-flat} steady states can appear.
They are called non-flat, because they can be represented in ($x$,$\phi$) 
as lines with non-zero slopes. 
Unlike flat steady states, lifetimes of these states
are finite in general depending on some parameters (will be discussed in detail later);
therefore, these states will be also called {\em metastable} states
(see Ref.\ \onlinecite{Benczik:2009} for another type of metastable states).
Note that non-flat states cannot be sustained in non-periodic cases 
due to the boundary effect\footnote{
If opinions at the boundary and those at the location right next to the boundary are 
not the same, opinions at the boundary will have a tendency to move toward the opinions at
the neighboring site. Therefore, for the whole state to be stable, the state has to be flat everywhere.}.

It is not hard to generalize the 1D results to 2D lattices, and 
%if we draw $\phi$ with respect to $x$ and $y$, 
steady states will be represented by 2D planes in $(x,y,\phi)$-space.
As in 1D cases, transient states will look like curved 2D surfaces mostly,
but eventually the system will reach one of steady states, however long it takes [see Fig.\ \ref{evolution}(b)].
%Since the lattice is finite in our model, boundary effects also exist, and 
When both the lattice and the opinion
are periodic, non-flat metastable states can emerge, while there will be only flat steady states otherwise.
When $g<1$, opinions at a location can have more than one value: in other words, at a given location there will be a distribution
of opinions.
For non-flat steady states, the width of the opinion distributions at a given location will be finite, 
and as $g$ gets smaller, the width of this distribution will increase.

% ======<Numerical Results>==========
\section{Numerical results for the periodic case}
\begin{figure}
\includegraphics[scale=0.5]{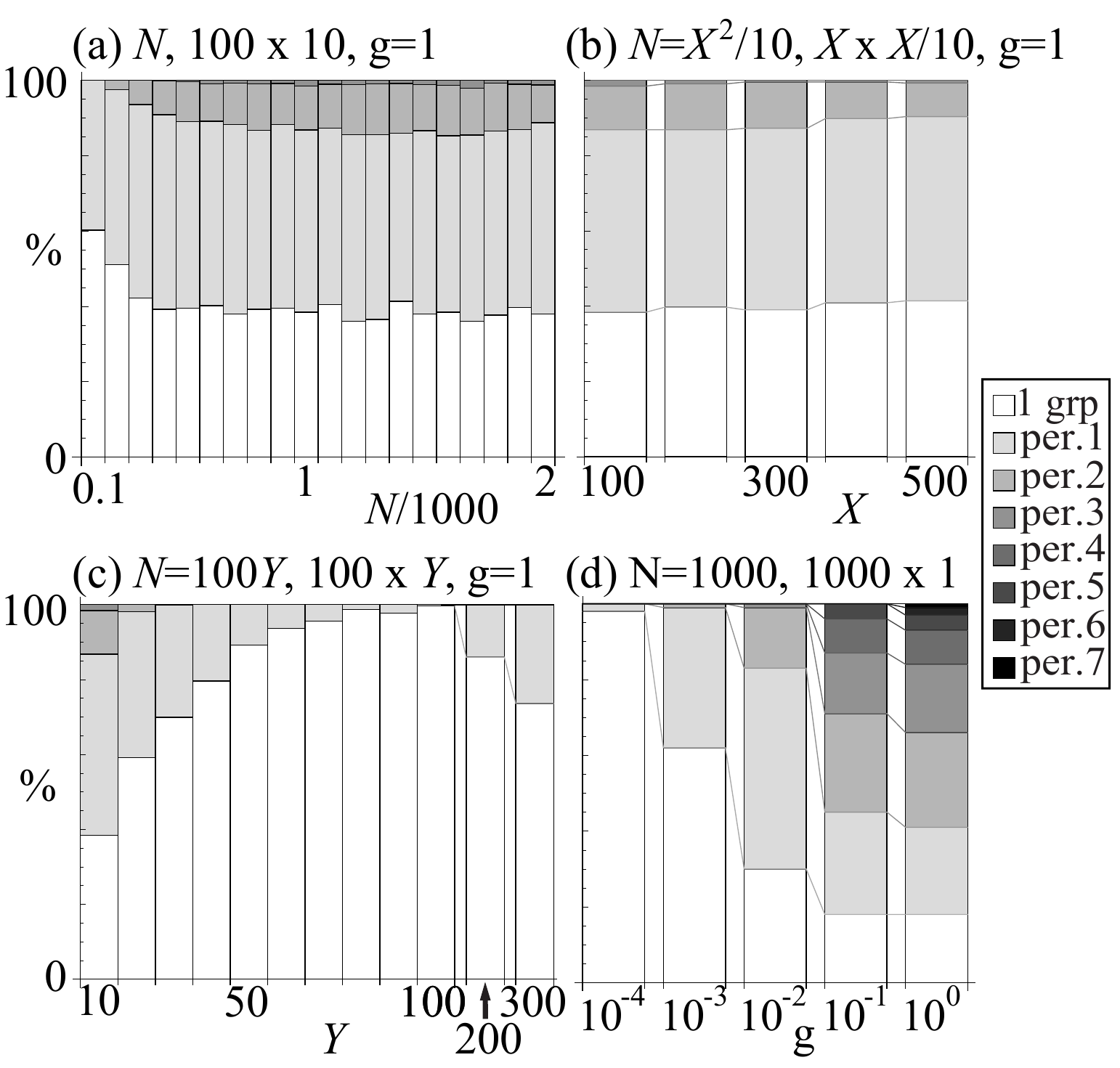}
\caption{\label{nxyg}
	Distributions of steady states varying $N$, $X$, $Y$, and $g$ ($d$ is fixed at 0.49, using 1000 runs for each case).
	(a) Varying $N$ ($X=100$, $Y=10$, $g=1$). The density changes.
	(b) Varying $N$, $X$ and $Y$ ($N=XY$, $Y=X/10$, $g=1$). The density and $Y/X$ do not change, and the size effect can be seen.
	(c) Varying $N$ and $Y$ ($N=XY$, $X=100$, $g=1$). The density and $X$ do not change, while the shape of the lattice changes.
	%We observe that when $Y/X$ gets closer to 1, periodic steady states almost disappear. 
	(d) Varying $g$ ($N=1000$, $X=1000$, $Y=1$).
	%As $g$ gets smaller, the number of periodic steady states also starts to decrease. 
	}
\end{figure}

%This model is mostly predictable, but numerical simulations are needed 
%to fully understand the dynamical behavior of the system in some cases.
%If we categorize our model based on structural properties, there are four cases: 
%rectangular lattices with non-periodic opinions, rectangular lattices with periodic opinions, 
%toroidal lattices with non-periodic opinions, and toroidal lattices with periodic opinions.
%For the first three cases, the dynamical behavior is mostly predictable. As we've seen in the previous section, steady states 
%will only be flat.
%If $d$ is big enough, the number of opinion group will be 1 always, but as $d$ gets smaller, there can be more than
%one opinion groups as we've also seen in other models\cite{BenNaim}.
The toroidal (periodic) lattice with periodic opinions
has richer dynamic behavior than other cases as we've seen in the previous  section.
In addition to the flat steady states, this system can have non-flat metastable states 
due to the periodicity of both the location and the opinion.
We can categorize these non-flat states with period numbers 
with ``period $n_\phi/n_l$" ($n_\phi,n_l=1,2,3,\ldots$; see Appendix for detail). 
For 2D cases, only one direction (either $x$ or $y$) can be periodic.
%, and
%if it is along the $x$ direction, for example, only when $\delta=(n_\phi/n_l)X$, states 
%will become steady states.
Even though any period-$(n_\phi/n_l)$ steady states can exist, we only observed steady states of period $n_\phi$ ($n_l=1$) and period $1/n_l$
($n_\phi=1$)
mostly in our numerical simulations, and we will show them in later figures.
For all numerical results except the one in Fig.\ \ref{nxyg}(a), we assume that the density of agents is 
always one per location.

\begin{figure}
\includegraphics[scale=0.4]{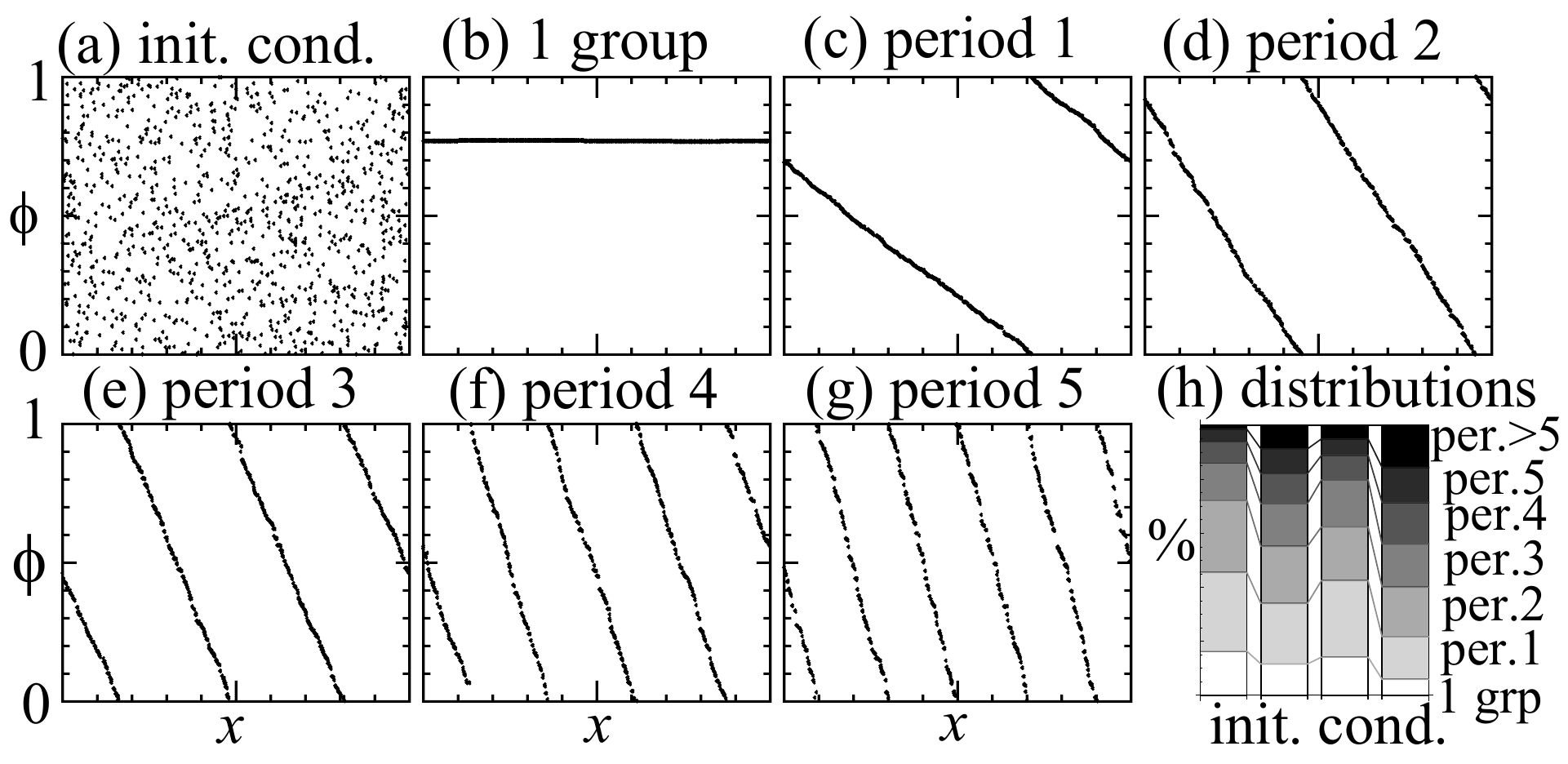}
\caption{\label{oneinit}
	Scatter plots for the case with
	1D periodic lattice and periodic 
	opinions ($N=1000$, $X=1000$, $Y=1$, $g=1$, and $d=0.49$), 
	(a) a randomly chosen initial condition;
	(b)-(g) observed steady states using the same initial condition in (a); 
	(h) a distribution of steady states using four randomly chosen initial conditions (using 1000 runs each). 
	}
\end{figure}

In Fig.\ \ref{oneinit}, 
%==== figure <oneinit>
%====
we observe 
steady states that can emerge for
$N=1000$, $X=1000$, $Y=1$, $g=1$, and $d=0.49$,
with one randomly chosen initial condition ($d=0.49$ is used as a maximal value
instead of 0.5 because 0.5 is excluded as was discussed in the previous section).
Since the dynamics is stochastic, the system
can reach the different types of steady states with the same initial condition.
In addition to the flat steady states [Fig.\ \ref{oneinit}(b)], non-flat steady states also appear
[Fig.\ \ref{oneinit}(c)-(g)].
By repeating simulations using different sets of random numbers, we can find the distribution of types of steady states
when we start from one given initial condition.
The distributions from different initial conditions don't have to be the same, as we've shown in Fig.\ \ref{oneinit}(h) for four
different randomly chosen initial conditions.
%For each initial condition, the system has been run 1000 times and the types of steady states have been counted.
%The more the number of runs, the more accurate these distributions will become. However, because we don't really need accurate distributions,
%1000 runs was enough to see differences.

How do we find out dynamical properties of the system with given parameters?
%One way is to find the distributions of types of steady states when we
%randomly choose samples of initial conditions and run simulations 
%for each initial condition the same number of times.
%Using just one initial condition repeatedly will not be enough to find out the overall dynamical behavior of the system
%as we've seen in Fig.\ \ref{oneinit}.
Each initial condition will have its own distribution 
as we've seen in Fig.\ \ref{oneinit}; therefore
the distribution we obtain after averaging over those from all initial conditions
characterizes the system.
%, which is a superposition of distributions for all possible initial conditions. 
The more samples we choose and the more runs we perform for each sample, the more accurate this distribution should be.
For each parameter set used subsequently, 
we will sample 1000 initial conditions randomly from the space of all initial conditions (in this case, $\{(x_i,y_i,\phi_i)|
1\le i\le N\}$),
and run once each, to find out the approximate distribution (the total of 1000 runs).

\begin{figure}
\includegraphics[scale=0.4]{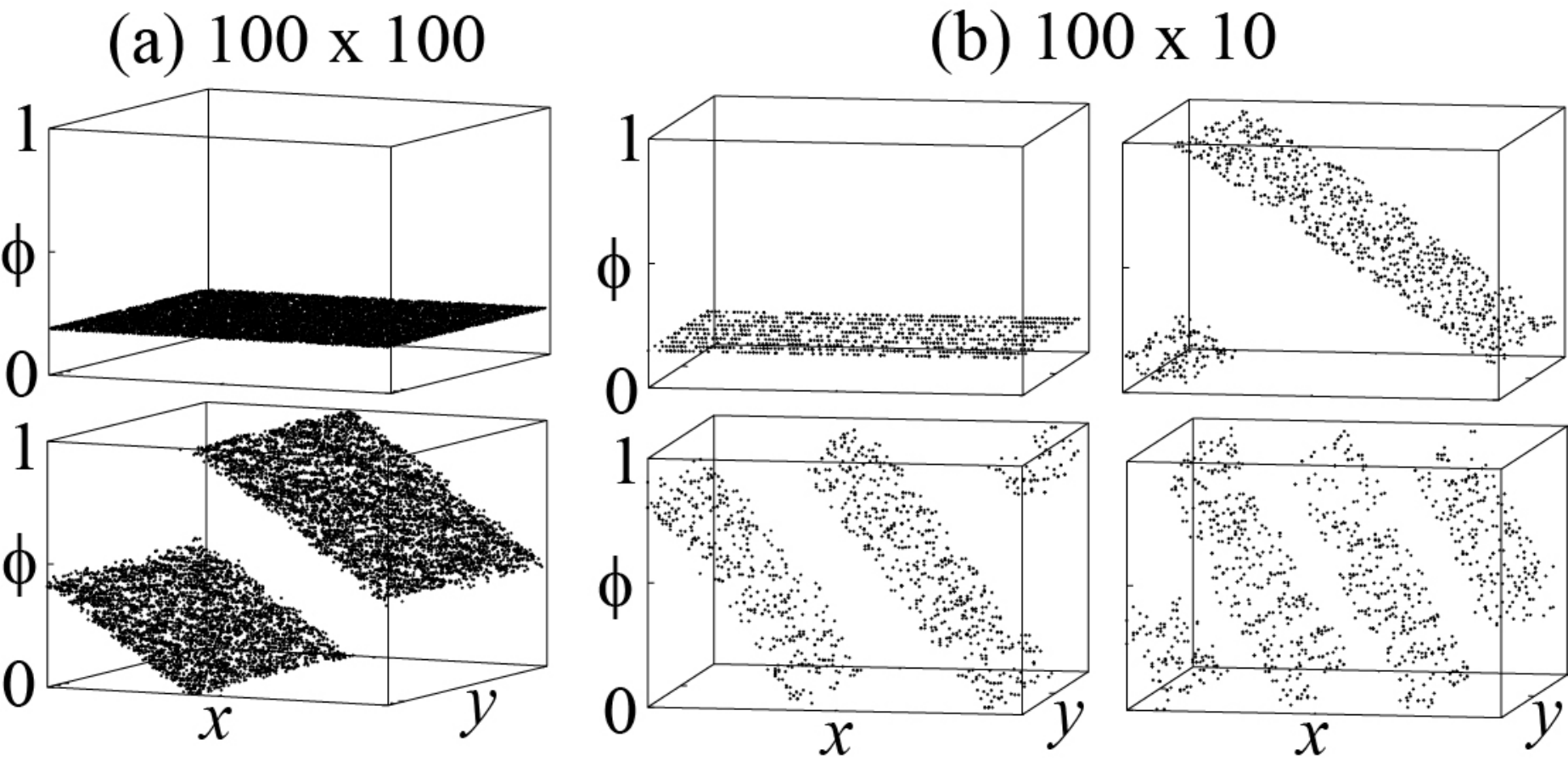}
\caption{\label{2dcase}
	Scatter plots in two cases with 2D periodic lattices 
	and periodic opinions.
%	There are more types of non-flat steady states when $Y$ is smaller 
%	compared to $X$.
	(a) $N=10\ 000$, $X=100$, $Y=100$, $g=1$, and $d=0.49$. 
%	Two types of steady states are observed: flat and period 1.
	(b) $N=1000$, $X=100$, $Y=10$, $g=1$, and $d=0.49$. 
%	Four types of steady states are observed:
%	flat, period 1, period 2, and period 3.
	}
\end{figure}

For 2D cases, the results are similar to those from 1D cases. In Fig.\ \ref{2dcase},
%==== figure <2dcase>
%====
we looked at steady states for two different cases: $(X,Y)=(100,100)$ and $(100,10)$. % for $N=10\ 000$ and $N=1000$, respectively.
In the first case, we observed two types of steady states: flat (\Tilde99\%) and period 1 (\Tilde1\%).
While, in the second case, we observed four types of steady states: flat (\Tilde 38\%), period 1 (\Tilde48\%), 
period 2 (\Tilde12\%), and period 3 (\Tilde2\%).
As we will show later in Fig.\ \ref{nxyg}(c), the width of the lattice 
in $y$-direction, $Y$,  can change the dynamical behavior of the system, when $X$ is fixed. 
When $X=Y$, the complete consensus was reached in almost all cases.

\begin{figure}
\includegraphics[scale=0.75]{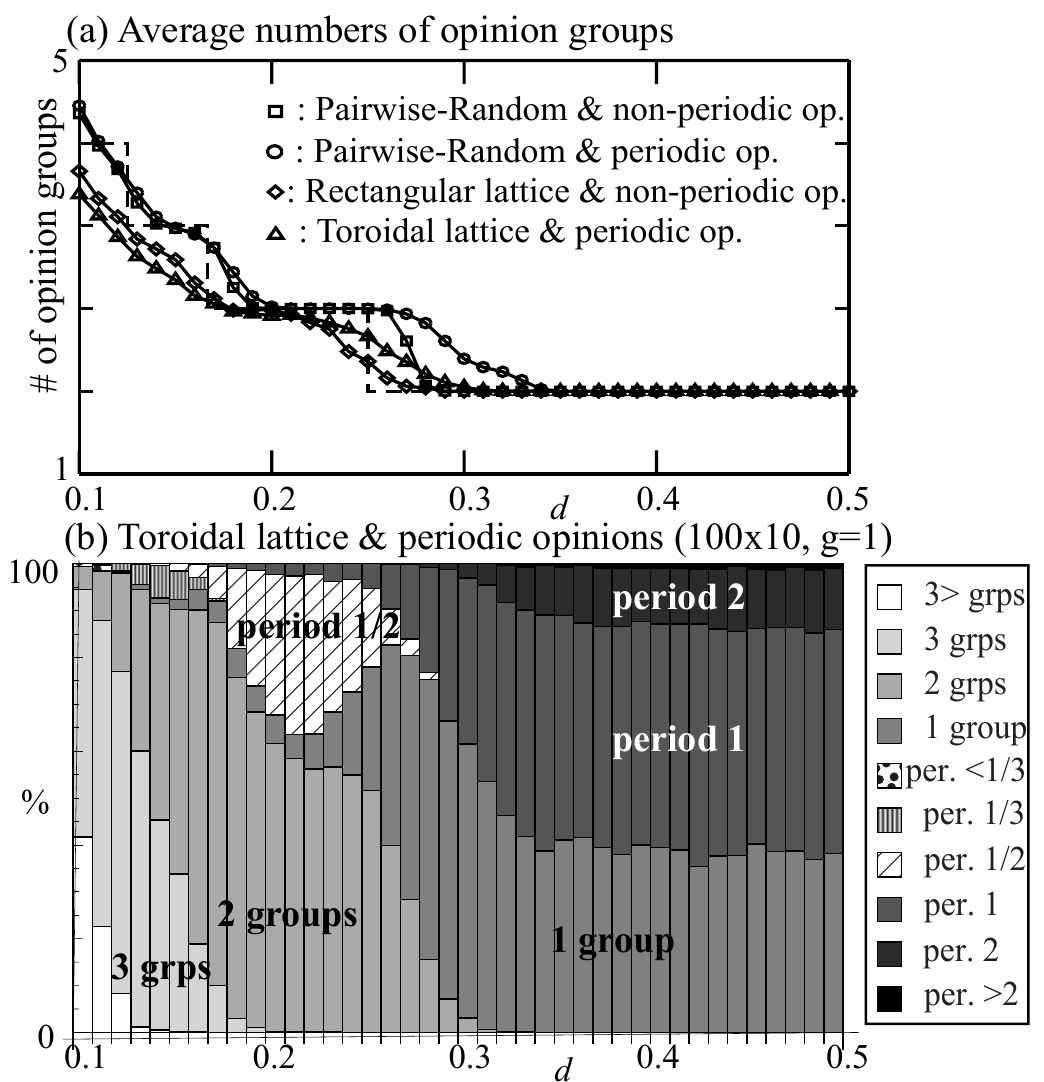}
\caption{\label{varyd}
	(a) Number of opinion groups formed when varying $d$ for four 
	different settings: 
	%pairwise-random (PR) interactions for both non-periodic and periodic opinions 
	%($N=1000$ and $g=1$, 1000 runs each), and 
	%nearest-neighbor (NN) interactions for the rectangular lattice with non-periodic 
	%opinions and the toroidal lattice with 
	%periodic opinions (
	$N=1000$, $X=100$, $Y=10$, $g=1$, 
	and $d=0.49$, 1000 runs each. 
	When counting numbers of groups, we counted groups with greater than 10\% of the whole population, and
	ignored non-flat steady states, if they exist.
	The dashed line represents the prediction that, when $d$ is in the range of $1/(n+1)<d<1/n$,
	there can be $n$ opinion groups.
	(b) Distributions of observed steady states for the case 
	with the toroidal lattice and periodic opinions, 
	while varying $d$ ($N=1000$, $X=100$, $Y=10$, $g=1$, 
	and $d=0.49$, 1000 runs each). 
	%Note that distributions for different parameter sets will be different from this result.
	}
\end{figure}

Let us next observe how the threshold $d$ changes the dynamic behavior of the system.
In Fig.\ \ref{varyd}(a), 
%==== figure <varyd>
%====
we compared our results with those from the Deffuant model by observing averaged numbers of opinion groups 
while varying $d$.
Two agents are randomly picked at each time step, and $g$ is set to  $1$.
%We can roughly predict that, when $d$ is in the range of $1/(n+1)<d<1/n$,
%there can be at most $n$ opinion groups in flat states.
In numerical results with $N=1000$ ($\Box$), averaged over 1000 runs, 
%numbers of groups were in line with the above prediction, and
gradual transitions were observed as the number of groups increases.
We simulated for periodic opinions ($\circ$), too, and got almost 
the same results. %, except a transition region from one group to two groups near $d=0.3$.
Two cases from our model were also simulated: a rectangular lattice with non-periodic opinions ($\Diamond$), and 
a toroidal lattice with periodic opinions ($\bigtriangleup$) when $N=1000$, $X=100$, $Y=10$, $g=1$, and $d=0.49$ for 1000 runs each.
The results were similar to those from the Deffuant model, 
but numbers of groups tend to be a little smaller.

In Fig.\ \ref{varyd}(b), steady states for the case of the toroidal lattice with periodic opinions 
with $N=1000$, $X=100$, $Y=10$, and $g=1$,
were observed when $d$ is varied.
There can be non-flat steady states. %as we've seen earlier, and we can observe what
%types of flat and non-flat steady states emerge when $d$ is given.
For flat steady states, we can categorize them with number of opinion groups: 1 group, 2 groups, 3 groups, and so on.
When $d$ is less than \Tilde0.17, steady states are mostly flat, and as $d$ gets smaller, the more opinion groups can exist
as seen in Fig.\ \ref{varyd}(a).
When $d$ is between \Tilde0.17 and \Tilde0.27, fractional periodic states, mostly 1/2, emerge, while dominant steady 
states are those with 2 groups.
When $d$ is between \Tilde0.27 and \Tilde0.32, this is where the transition occurs: the number of 2-group steady states 
decreases quickly, while the number of period-1 steady states increases.
When $d$ is greater than \Tilde0.32, all flat steady states belong to the 1-group type, while there can be many types of
non-flat steady states. 
Distributions don't change much as $d$ increases up to 0.5.
In the current case, steady states with period 3 or higher don't appear much; however when $Y/X$ is smaller, more types 
of periodic steady states will appear.

\begin{figure}
\includegraphics[scale=0.37]{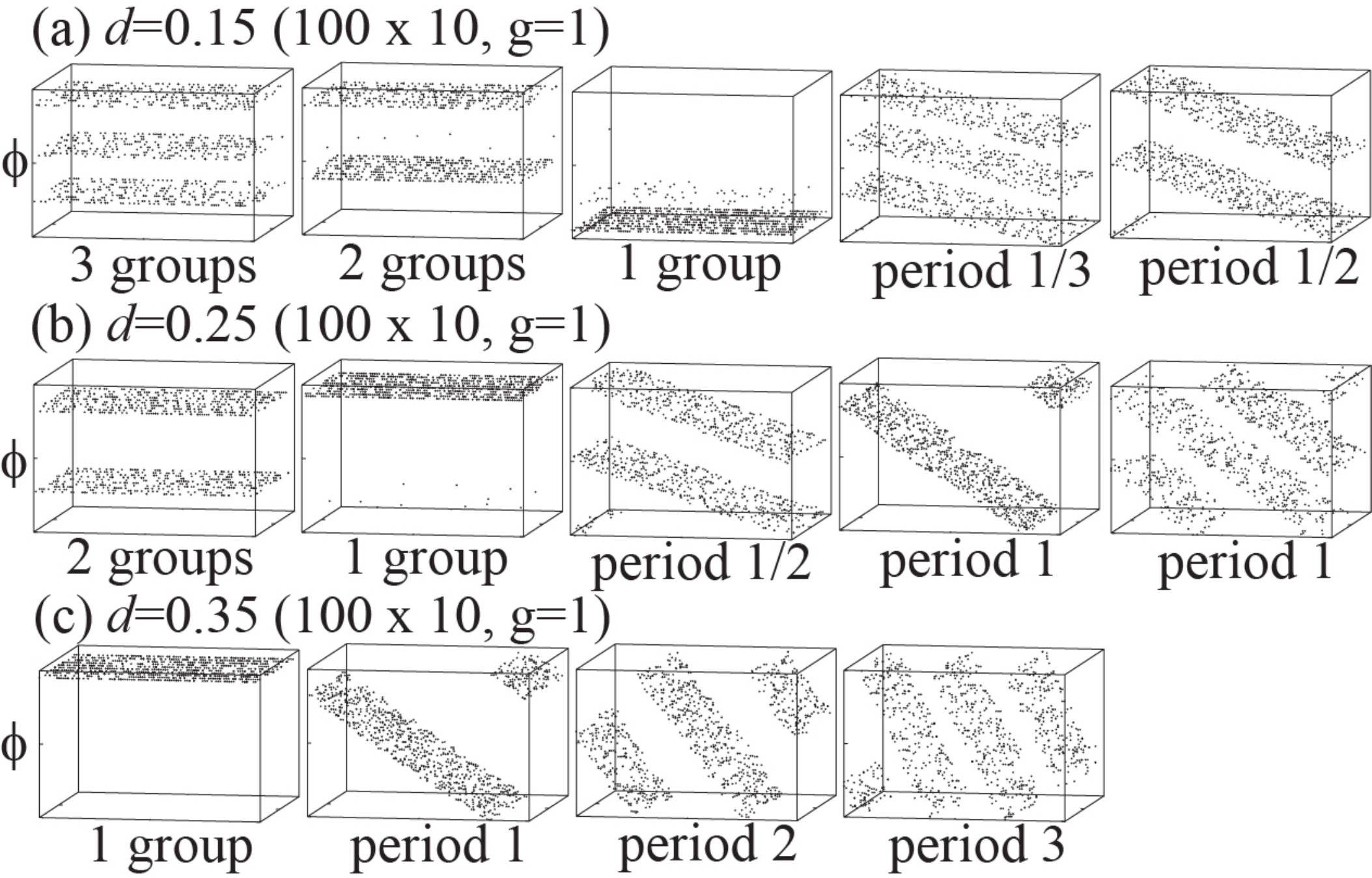}
\caption{\label{observed}
	Scatter plots of some observed steady states for different $d$ values 
	from Fig.\ \ref{varyd}(b) 
	($N=1000$, $X=100$, $Y=10$, $g=1$, and $d=0.49$, 
	toroidal lattice with periodic opinions). 
	(a) $d=0.15$ (3 groups: 34\%, 2 groups: 56\%, 1 group: 2\%, period 1/3: 6\%, period 1/2: 2\%),
	(b) $d=0.25$ (2 groups: 52\%, 1 group: 26\%, period 1/2: 17\%, period 1: 5\%, period 1: $<1$\%),
	(c) $d=0.35$ (1 group: 41\%, period 1: 48\%, period 2: 10\%, period 3: 1\%). 
	}
\end{figure}

In Fig.\ \ref{observed},
%==== figure <observed>
%====
we show steady states in Fig.\ \ref{varyd}(b) for three $d$ values: 0.15, 0.25, and 0.35..
When $d$ is 0.15, three types of flat steady states dominate (\Tilde92\%), while there exist non-flat steady states with
fractional periods (mostly 1/2 and 1/3).
When $d$ is 0.25, more than 3/4 of steady states are flat (\Tilde78\%) still, while non-flat steady states with period
1/2 and 1 also exist. Here there are two kinds of period-1 steady states: one has one band ($n_\phi=1$ and $n_l=1$), 
and the other has two bands ($n_\phi=2$ and $n_l=2$),
which is possible because $d$ is small. 
When there are two bands, there are two disjoint groups of agents,
% in which only agents of the same group interact with each other 
even though each group has agents with a full spectrum of opinions, and
they were observed only in the approximate range of $d$ between 0.18 and 0.25.
When $d$ is 0.35, only 1-group flat states were observed, and non-flat steady states of periods 1, 2, and 3 were also observed.

Finally, we can ask how other parameters will influence the outcome.
In Fig.\ \ref{nxyg}(a),
%==== figure <nxyg>
%====
$N$ varies from 100 to 2000 when $X=100$, $Y=10$, $g=1$, and $d=0.49$
to observe how the density [$N/(XY)$] changes dynamical behavior.
When the density is much smaller than 1, the number of non-flat steady states decreases.
But we can clearly observe that if the density is greater than about 0.5, distributions do not change much.
This result shows that the density doesn't have to be high to find out distributions of steady states,
and that the density of 1 is good enough.

In Fig.\ \ref{nxyg}(b), we vary 
the size of the lattice while keeping the density and the shape fixed.
As $X$ increases from 100 to 500 when $N=XY$, $Y=X/10$, $g=1$, and $d=0.49$, the distributions do not change much.
We can interpret this result as a sign that there is no size effect unless the size is too small.

In Fig.\ \ref{nxyg}(c),
$Y$ varies from 10 to 300, changing the shape of the lattice, when $N=XY$, $X=100$, 
$g=1$, and $d=0.49$.
As $Y$ approaches the value of $X$, 
the number of non-flat steady states decreases.
As we've seen in Figs.\ \ref{2dcase} and \ref{observed}, the periodic behavior 
only appears in one direction in the case of toroidal shapes.
If opinions are periodic in $x$, opinions along the $y$ direction for a given $x$ 
value are more or less constant.
%Then, the wider $Y$ gets, the harder it is for agents to form periodic steady 
%states and the less likely these steady states are reached.
When $Y$ is 10, periodic behavior can be only seen in $x$, but as $Y$ 
increases, non-flat steady states along the $y$ direction starts to emerge. 
When $X=Y$, steady states can be periodic either in $x$
or in $y$ with equal probabilities.
When $Y$ is greater than $X$,
the likelihood of forming steady states that are periodic in $y$ will be greater,
and  the probability of getting the periodic steady states starts to increase again.
%Even for 1D cases ($Y=1$), as the size $X$ increases, 
%the likelihood of reaching steady states of higher periods becomes greater.
The more elongated the shape of the lattice, the greater is the
likelihood of finding periodic steady states.

In Fig.\ \ref{nxyg}(d), the convergence parameter
$g$ varies when $N=1000$, $X=1000$, $Y=1$, and $d=0.49$.
As $g$ gets smaller, the number of non-flat steady states decreases.
%Here there are no steady states with fractional periods, because $d$ is maximal. 
To make those periodic steady states disappear, 
$g$ has to be very small, \Tilde$10^{-4}$, in this case.
%What are the effects of $g$ then?
%Here $g$ is the coupling constant, the strength of the interaction. When $g$ is 
%equal to 1, opinions of 
%agents at the same location converges at the next time step; on the other hand, 
%the smaller $g$ gets,
%the smaller the amount of change of opinions, which means that it takes longer 
%for opinions to converge.
The convergence parameter controls how fast opinions converge;
in addition, when $g$ is small, the distribution of opinions at the same location 
for non-flat steady states gets wider
because agents can move farther away from a location without changing their 
opinions much.

\begin{figure*}
\includegraphics[scale=0.7]{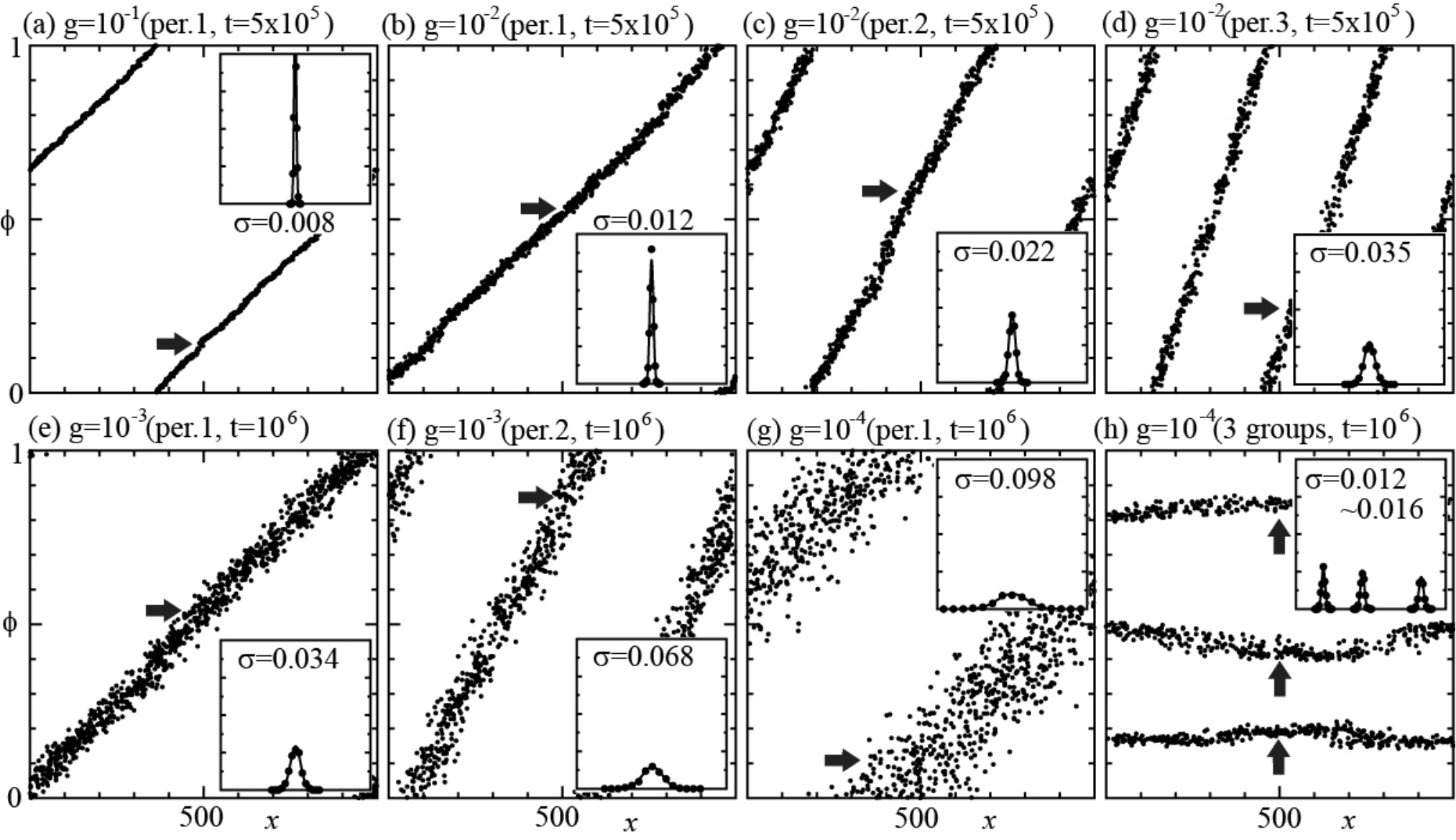}
\caption{\label{width}
Scatter plots of non-flat steady states from Fig.\ \ref{nxyg}(d).
%We can find widths of opinion distributions at fixed $x$ values.
The widths were obtained by fitting the distributions
of $\phi$ at $x=500$ (pointed by arrows) by Gaussian functions, and 
$\sigma$ is the standard deviation (shown in insets).
	}
\end{figure*}

In Fig.\ \ref{width},
%==== figure <width>
%====
we observe widths of distributions of opinions at a given $x$ for some of 
non-flat steady states from Fig.\ \ref{nxyg}(d).
In Figs.\ \ref{width}(a)-(g), non-flat steady states when $g=10^{-1}$, 
$10^{-2}$, $10^{-3}$, and $10^{-4}$ are shown.
Insets show distributions of opinions $\phi$ at $x=500$.
%Here frequencies of opinions were obtained by counting the opinions 
%from states at a series of time steps after steady states had been reached.
%At each time step, there will be an average of one agent at $x=500$
%in this case, and,
%to get the frequency distributions, that is not enough; hence,
We use opinions of agents residing at $x=500$ at
a number of different time steps after steady states are reached  
(for example, 1000 time steps with 100 time steps apart)\footnote{
The width found this way can be a little bigger than the actual width,
because the band structures in $(x,\phi)$-graphs tend to drift
a little in the $x$ direction even after the steady state has been reached.
This effect has been ignored here.
}. 
If these frequencies are fitted to Gaussian functions as shown in insets in Fig.\ 
\ref{width}, 
the standard deviations, $\sigma$,
will represent the widths of these distributions.
We can argue that $\sigma$ is a measure of stability.
%How is $\sigma$ related to stabilities of certain types of states? 
%First, it's necessary to define what the stability means here.
As discussed in Sec.~II,
we can call non-flat steady states {\em metastable}, because lifetimes of structures
formed in $(x,y,\phi)$-space are finite, but extremely long.
%When a periodic state is stable, 
%we've been calling  the state a steady state, and this steady state 
%does not morph into another type of steady states in simulations. 
In other words, the probability of decay within a certain finite time period
is extremely small, even though it will not be zero exactly as in flat steady states.
%Here $\sigma$ is determined by given parameters and the period number $n_\phi/n_l$,
If we define $R$ as
the ratio of the width of a band
to the distance between two adjacent bands along the $x$ direction,
\begin{equation}
R=\frac{(n_lX/n_\phi)\sigma}{X/n_\phi}=n_l\sigma,
\end{equation}
where $n_lX/n_\phi$ is the inverse of the slope of the line in $(x,\phi)$.
Since $n_l=1$ always in current examples, $\sigma=R$.
If $R\ll0.5$, the structure is quite stable.
%[in Fig.\ \ref{width}(g), we can observe
%that states were stable when $R=\sigma\simeq0.1$].
If $R>\Tilde0.5$, adjacent bands will overlap and
will be broken immediately.
If $R$ comes close to 0.5, the state
is likely to become flat or morph into another type of steady states in a short time period. 
In general, the bigger $\sigma$ is, the more unstable the state is and the less likely
it will be reached.

How is $\sigma$ affected by parameters?
We have observed in Fig.\ \ref{width} that the greater $g$, the smaller is $\sigma$.
This width also depends on the period number $n_\phi/n_l$; for example, the width of 
opinions of period-2 steady states
is twice as wide as that of period-1 steady states, 
because the opinion difference between two adjacent locations is twice as 
big for period-2 steady states.
Then we can generalize that for the same $g$ value, the width of opinions at a 
location for steady states with period $n_\phi/n_l$
is proportional to $n_\phi/n_l$. 
This explains why steady states with higher periods disappear quickly as $g$ 
gets smaller.
When $g=1$, $\sigma$ comes close to zero unless $N$ is very small.
That leads us to claim that $N$ can play a role, too:
as $N$ gets smaller, $\sigma$ becomes wider because,
when $N$ is small compared to $XY$, the possibility of having empty locations increases
and an agent can move further away without changing its opinion
[see Fig.\ \ref{nxyg}(a)].
In addition, stability of non-flat states also depends on $Y/X$
[see Fig.\ \ref{nxyg}(c)].
We can summarize based on our results so far that, in general, 
the stability of a certain periodic state depends directly on 
$N/(XY)$, $Y/X$, $g$, $n_\phi$ and $n_l$, while $d$ determines what types 
of periodic states are allowed to exist.

Figure \ref{width}(h) shows a special case.
It has three opinion groups, but unlike opinion groups found in 
flat states when $d$ is small, 
they have non-zero widths of opinions and an opinion
of each agent is not stationary.
In addition, the lifetimes of these states are not long.
In most cases, this type of states stayed intact for more than 10 times of that
it took for most steady states to be formed.
This type of states were observed only in the case with $g=10^{-4}$
for about one percent [note that since they eventually converge to flat states,
they were counted as flat states in Fig.\ \ref{nxyg}(d)].
They could emerge, because periodic opinions are used and
opinions from three groups can be balanced in a circle.
Also, $g$ has to be small enough; if $g$ is bigger, widths of opinions
will be bigger and the structure becomes unstable.

%=======<Discussions>=============
\section{Discussions}
We introduced a simple model of opinion dynamics that uses only nearest-neighbor interactions.
To represent the geometrical space we live in, we used 2D lattices, on which agents move randomly.
Both the lattice and the opinion can be periodic or non-periodic.
We explored some regions of the parameter space, and found rich
dynamic behavior, especially when both are periodic.
One might argue that periodic opinions and toroidal lattices are not realistic and even artificial.
But the opposite might be true: linear opinions and rectangular lattices are rather special cases.
The surface of the Earth is finite and has no boundary.
Opinions, discrete or continuous, are not always linear, either.
In some cases, opinions can be better represented by more general structures other than a line.
In short, our results show that if we generalize spaces for opinions and lattices, 
dynamic behavior of those systems can be richer.

In reality, a consensus is not reached easily.
Consider political opinions as an example.
Even though political parties have been formed in advanced societies,
people in one party usually have various political views.
Another example is the existence of dialects, if we regard languages as opinions.
%Usually, agents have the wide
%spectrum of opinions in a society even though we interact with others and are influenced by them.
%There are several explanations for this coexistence of all possible opinions:
These phenomena can be explained by inherent heterogeneity of agents, or 
bounded confidence.
Our model, however, adds another explanation: metastability through {\em local} interactions.
Metastable states in our model indicates that
%the local convergence does not guarantee the global convergence (a complete consensus).
locally-converged, but not globally-converged, states 
can be sustained for a long time in certain situations.
For example, if we are surrounded by like-minded people, we seldom change
our views and even believe that everybody is similar to us, 
which isn't true in general.
%, if agents are interacting with only nearest neighbors,
%the process of repeated averaging may not lead to steady states representing consensus.
%Nearest-neighbor interactions with random walks are used to mimic this phenomenon.
%In addition, most of us don't change our political views easily, 
%which can be realized in our model with small $g$ values.

In general, steady states or equilibrium states in models that are 
closed are hard to be realized in real systems like 
societies because societies are fundamentally open and extremely noisy. 
However, behavior of transient states 
and emergence of different types of steady states can shed some light on 
understanding how real systems behave.
In this model, metastable states, which have the full spectrum of opinions,
are only observed in cases of toroidal lattices with periodic opinions.
But even for cases with rectangular lattices,
as the size of the lattice gets bigger,
the overall behavior of locally-converged transient states seen in Fig.\ \ref{evolution}
will be similar as long as the opinions are periodic.
Then we can interpret our real-life states containing the wide spectrum of opinions
as transient states that are moving slowly toward steady states.

The model considered here can be expanded or modified.
We can assume $g$ and $d$ are not constant throughout the whole population
(so called, {\em heterogeneous} agents\cite{BenNaim:2003,Weisbuch:2002,Iniguez:2009}).
%Then the model will mimic real systems better, because in reality people are heterogeneous:
%some people are more open to others' ideas than others (bigger $d$), and some people have a greater tendency
%to be influenced by others (bigger $g$).
%We expect that it can change the dynamical behaviors of the system in some cases.
The definition of neighbors can be modified by using
a bigger range, so that the network structure becomes more realistic.
%In reality, agents in a society change continuously; hence we can incorporate birth and death of agents
%in the model, thereby making the model {\em open}, which can also change some dynamical behaviors.
%Opinions are continuous here, but we can devise similar models with
%discrete opinions, where the number of opinions is not small.
%Then the system becomes a Markov chain with discrete states,
%which can be represented by a transition matrix.
%This Markov chain will not be irreducible in models that are similar to our model, 
%and complicated limiting behavior as we've seen in our model is also expected. 
%On the other hand, we can use a variety of interaction rules on discrete opinions,
%where the number of opinions is not big (for example, the voter model\cite{Liggett}),
%and study their dynamical behaviors.
%this model is to investigate models with the dynamics of social networks.
%Even though it wasn't mentioned here, nearest-neighbor relations
%can be regarded as an evolving network that guides changes of agents' opinions.
%One can compare our model with other models that use nearest-neighbor interactions using 
%fixed neighbors like the voter model in cellular automata\cite{Durrett}.
%In those models, the network structure that defines neighbors does not change, while,
%in our model, the network evolves with time, which resembles the reality better.
We can also consider an additional co-evolving network like an acquaintance network on top of our model,
%an acquaintance network that also co-evolve with other agents.
%it will be interesting to see what dynamical behaviors can emerge from such systems.
and it will be investigated in a future article.

The models for social dynamics like this are not devised for predicting the future of specific systems, but
can help us understand dynamical properties.
In addition, they can be adapted to a wide variety of different systems with similar features.
%We hope this is one step toward the goal of understanding dynamical behaviors of seemingly too-complicated systems.

\begin{acknowledgments}
The author thanks Chang-Yong Lee for his help on finding out that 
some unexpected results were actually metastable states,
while at first it had seemed to be from unknown numerical errors.
The author also thanks L.\ E.\ Reichl for her thoughtful comments 
concerning this manuscript.
This work was supported by Kongju National University.
\end{acknowledgments}

\appendix*
\section{Solving an 1D case}
We make the lattice one-dimensional by setting $Y=1$, and 
we also assume $g=1$ and the maximal $d$ ($d=1$ for non-periodic opinions, and $d=0.49$ for periodic opinions).
If the density of agents, the average number of agents at each location, is high enough,
we can simplify the system by making the opinion to be the function of the location, 
ignoring the states of individual agents.
That is due to the fact that every agent in the same location will have the same opinion instantly.
%because $g=1$ and  $d$ is maximal.
Then the overall $N$-agent dynamics is reduced to the dynamics of $\phi(x)$, an $X$-dimensional vector.

\begin{figure}
\includegraphics[scale=0.7]{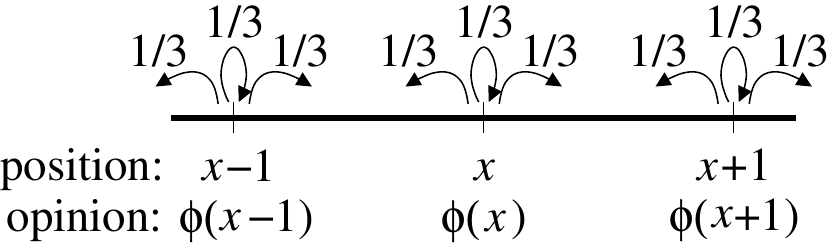}
\caption{\label{explanation}
	Schematic diagram for an 1D case ($Y=1$) with $g=1$ and the maximal $d$.
%	Because there are only 3 possible moves, each move has the probability 1/3.
	}
\end{figure}

Since an agent has three choices to move (left, right, and staying) with the probability 1/3 for each move
(see Fig.\ \ref{explanation}),
%==== figure <explanation>
%====
the opinion at $x$, $\phi(x)$, at the next time step will become $[\phi(x-1)+\phi(x)+\phi(x+1)]/3$.
%because a third of the population
%at $x$ will come from $x-1$, another third from $x+1$, and the other third from the same location, $x$.
This is a discrete-time linear dynamical system, which can be characterized by
the tridiagonal matrix with all non-zero elements 1/3.
Usually we are interested in finding states where $\phi(x)$ does not change with time, and 
they are eigenstates of this matrix with the eigenvalue 1.
If we look at the dynamics locally,
when $\phi(x-1)=\phi(x)-\delta$ and $\phi(x+1)=\phi(x)+\delta$, $\phi(x)$ will not change at the next time step,
where $\delta$ is a small real number at $x$.

Since both $x$ and $\phi$ are bounded, we cannot find eigenstates easily, except one trivial case 
when $\phi(x)$ is constant with respect to $x$ ($\delta=0$ for all $x$).
If a system arrives at these states, a consensus has been reached.
In most cases, these are the only eigenstates, but when both the lattice and the opinion are periodic,
the periodic eigenstates can exist as long as the boundary conditions are met, where $\delta$
is a non-zero constant for all $x$.
However, $\delta$ needs to be certain discrete values,
and we can name these periodic eigenstates with period numbers.
The ``period $n_\phi/n_l$" ($n_\phi,n_l=1,2,3,\ldots$) 
means that opinions have $n_\phi$ cycles of changes while there are $n_l$ cycles of changes along the $x$ direction;
hence only when $\delta=(n_\phi/n_l)X$, they will become eigenstates.

\bibliography{p12_e1}%{}

\end{document}